\documentstyle[12pt]{article}

\newcommand{\alt}{\mathrel{\raisebox{-.6ex}{$\stackrel{\textstyle<}{\sim}$}}}
\newcommand{\agt}{\mathrel{\raisebox{-.6ex}{$\stackrel{\textstyle>}{\sim}$}}}

\def\overlay#1#2{\ifmmode \setbox 0=\hbox {$#1$}\setbox 1=\hbox to\wd 0{\hss
$#2$\hss }\else \setbox 0=\hbox {#1}\setbox 1=\hbox to\wd 0{\hss #2\hss }\fi
#1\hskip -\wd 0\box 1}

\catcode`@=11
\newcount\@tempcntc
\def\@citex[#1]#2{\if@filesw\immediate\write\@auxout{\string\citation{#2}}\fi
  \@tempcnta\z@\@tempcntb\m@ne\def\@citea{}\@cite{\@for\@citeb:=#2\do
    {\@ifundefined
       {b@\@citeb}{\@citeo\@tempcntb\m@ne\@citea\def\@citea{,}{\bf ?}\@warning
       {Citation `\@citeb' on page \thepage \space undefined}}%
    {\setbox\z@\hbox{\global\@tempcntc0\csname b@\@citeb\endcsname\relax}%
     \ifnum\@tempcntc=\z@ \@citeo\@tempcntb\m@ne
       \@citea\def\@citea{,}\hbox{\csname b@\@citeb\endcsname}%
     \else
      \advance\@tempcntb\@ne
      \ifnum\@tempcntb=\@tempcntc
      \else\advance\@tempcntb\m@ne\@citeo
      \@tempcnta\@tempcntc\@tempcntb\@tempcntc\fi\fi}}\@citeo}{#1}}
\def\@citeo{\ifnum\@tempcnta>\@tempcntb\else\@citea\def\@citea{,}%
  \ifnum\@tempcnta=\@tempcntb\the\@tempcnta\else
   {\advance\@tempcnta\@ne\ifnum\@tempcnta=\@tempcntb \else \def\@citea{--}\fi
    \advance\@tempcnta\m@ne\the\@tempcnta\@citea\the\@tempcntb}\fi\fi}
\catcode`@=12

\begin{document}

\font\fortssbx=cmssbx10 scaled \magstep2
\hbox to \hsize{
\hskip.5in \raise.1in\hbox{\fortssbx University of Wisconsin - Madison}
\hfill\vtop{\baselineskip16pt
            \hbox{\bf MADPH-95-865}
            \hbox{\bf CPP-95-1}
            \hbox{\bf DOE-ER40757-061}
            \hbox{\bf UCD-95-2}
            \hbox{\bf RAL-95-003}
            \hbox{\bf hep-ph/9501379}
            \hbox{January 1995}
                }}

\vspace{.3in}

\begin{center}
{\large\bf Probing Strongly-interacting Electroweak Dynamics through {\boldmath
$W^+W^-/ZZ$} Ratios at  Future \boldmath{$e^+e^-$}  Colliders}\\[.2in]
V.~Barger$^a$, Kingman Cheung$^b$, T.~Han$^c$ and R.J.N.~Phillips$^d$\\[.1in]
{\it
$^a$Physics Department, University of Wisconsin, Madison, WI 53706, USA\\
$^b$University of Texas at Austin, Center for Particle Physics, Austin, TX
78712, USA\\
$^c$Physics Department, University of California, Davis,
   CA 95616, USA\\
$^d$Rutherford Appleton Laboratory, Chilton, Didcot, Oxon OX11 0QX, UK}
\end{center}

\vspace{.2in}
\begin{abstract}
We point out that the ratio of $W^+ W^- \to W^+ W^-$ and $W^+W^- \to ZZ$
cross sections is a sensitive probe of the dynamics of electroweak
symmetry breaking, in the CM energy region $\sqrt {s_{\rm ww}^{}} \agt 1$~TeV
where
vector boson scattering may well become strong.  We suggest ways in
which this ratio can be extracted at a 1.5 TeV $e^+e^-$  linear collider,
using  $W^\pm,Z \to jj$  hadronic decays and relying on dijet mass
resolution to provide statistical discrimination between $W^\pm$ and $Z$.
$WW$ fusion processes studied here  are unique
for exploring scalar resonances of mass about 1 TeV
and are complementary to  studies via the direct
channel  $e^+ e^- \rightarrow W^+W^-$ for the vector and non-resonant cases.
With an integrated luminosity of 200 fb$^{-1}$,
the signals obtained  are statistically significant.
Comparison  with a study of $e^-e^- \rightarrow \nu\nu W^-W^-$ process
is  made. Enhancements of the signal rate from using a polarized electron
beam, or  at a 2 TeV $e^+e^-$ linear collider and
possible higher energy $\mu^+\mu^-$ colliders, are also presented.
\end{abstract}

\thispagestyle{empty}
\newpage

\section{Introduction}

The mechanism of electroweak symmetry breaking (EWSB) is the foremost
open question in particle physics today.  One direct approach to this
question is to search for Higgs bosons~\cite{hhg}.  A complementary
approach is to study the scattering of pairs of longitudinally polarized
 weak bosons \cite{reviews,hantalk,bagger}
$W_L$ (where and henceforth, $W$ generically denotes $W^\pm$ and
$Z$ unless specified otherwise),
since at high energies they recall their origins
as Goldstone bosons and reflect the EWSB dynamics, thanks to an equivalence
theorem \cite{gbet}.  In the Standard Model (SM), if the Higgs
boson is not very heavy $(m_H \alt 0.5$~TeV), $W_L W_L$ scattering
remains  relatively weak.  But in general, if there is no Higgs boson
below about 0.8 TeV,
the scattering of $W_L$  pairs is expected to become strong at CM
energies of order $\sqrt {s_{\rm ww}^{}} \agt 1$~TeV.   A variety of models
of the Strongly-interacting  Electro-Weak Sector (SEWS) have been
put forward to parameterize this strong scattering, to impose the
constraints of unitarity and crossing, and to characterize different
EWSB possibilities~\cite{reviews,hantalk,bagger}.

In the present paper we first point out that the cross section ratio
$\sigma(W^+ W^- \to W^+ W^-)/$ $\sigma (W^+W^- \to ZZ)$ is a sensitive
probe of the SEWS, since different models predict very different
ratios.  We then suggest ways in which this ratio may be
extracted from a  ``Next  $e^+e^-$ Linear Collider'' (NLC)
with the  CM energy $\sqrt s=1.5$~TeV  through the $W^+W^-$
fusion processes~\cite{fusionjfg,hagi,fusionvb,najima}
\begin{equation}
e^+ e^-
\to \bar\nu \nu W^+W^- ,  \bar\nu \nu ZZ \;.
\label{eq:WWfusion}
\end{equation}
In studies of strong $WW$ scattering at hadron colliders, it is necessary for
identification to use leptonic $W^\pm,Z$ decays, which have the disadvantages
of an invisible neutrino and/or small branching fractions. At $e^+ e^-$
colliders we are able to exploit the hadronic decays, which have the
advantages of large branching fractions and reconstructibility.
Here we rely on $W^\pm, Z \to jj$  hadronic decays, with sufficient
dijet mass resolution to provide statistical discrimination between
$W^+W^-$ and $ZZ$ final states.
We suggest cuts to minimize the principal
backgrounds from transverse $W$-pair production that are intrinsic in
Eq.~(\ref{eq:WWfusion}), and also come from
\begin{equation}
   e^+ e^- \to e^+ e^-W^+W^-, e^+ e^-ZZ, e^{\pm}\nu W^{\mp}Z
\,, \label{eq:bkgds}
\end{equation}
where the final-state electrons escape undetected along the beam-pipe,
as well as from the annihilation channel~\cite{annihil}
\begin{equation}
   e^+ e^- \to ZW^+W^- \to \bar \nu\nu W^+W^-     \,.
\end{equation}
We discuss the prospects for discriminating between $W^+W^-, \  W^\pm Z$
and $ZZ$ final states and make illustrative calculations to show what
may be learned from experiments.

The process  $e^-e^- \rightarrow \nu\nu W^-W^-$
is  unique to explore the weak ``isospin'' $I=2$
non-resonant channel \cite{hantalk,bbch}.
We therefore include a comparison of results at both $e^+e^-$ and
$e^-e^-$ colliders.
We also show the improvements that would come from
using polarized electron beams,  in both $e^+e^-$ and $e^-e^-$ cases.

The SEWS effects become significantly larger as energy increases.
We therefore demonstrate the enhancement of the signal rate at
a 2 TeV $e^+e^-$ collider and
possible $\mu^+\mu^-$ circular colliders with larger CM
energies~\cite{sausalito}.

The paper is organized as follows:
Section II presents the models to be compared.  Section III discusses
the question of dijet mass resolution.  Section IV describes
our methods of calculation and the motivation for our choices of acceptance
cuts. Results are presented and discussed in the final two sections.
We conclude that  at  a 1.5 TeV NLC with an integrated luminosity
of 200 fb$^{-1}$, it should be feasible to extract information on SEWS
by separately studying $W^+W^-$ and $ZZ$ events.

\section{Models for $W_LW_L$ scattering}

If we ignore gauge couplings and the mass $M_W^{}$ the scattering of
real longitudinal weak bosons  $W_L^1 W_L^2 \to W_L^3 W_L^4$
due to
EWSB interactions is the same as the scattering of the corresponding
Goldstone bosons \cite{gbet}
and can be parametrized by an amplitude $A(s,t,u)$ as follows:
\begin{eqnarray}
M(W_L^+W_L^-\to Z_LZ_L) &=& A(s,t,u),\\
M(W_L^+W_L^-\to W_L^+W_L^-) &=& A(s,t,u) + A(t,s,u),\\
M(Z_LZ_L\to Z_LZ_L) &=& A(s,t,u)+A(t,s,u) + A(u,t,s),\\
M(W_L^{\pm}Z_L\to W_L^{\pm}Z_L) &=& A(t,s,u),\\
M(W_L^{\pm} W_L^{\pm} \to W_L^{\pm} W_L^{\pm}) &=& A(t,s,u) + A(u,t,s),
\end{eqnarray}
where $s=(p_1 + p_2)^2$, $t=(p_1 - p_3)^2$ and $u=(p_1 - p_4)^2$ are
the usual Mandelstam variables.
We recall that the amplitudes $T(I)$
for total isospin $I$, which should obey unitarity, are then given by
\begin{eqnarray}
T(0) &=& 3A(s,t,u) + A(t,s,u) + A(u,t,s),\\
T(1) &=&             A(t,s,u) - A(u,t,s),\\
T(2) &=&             A(t,s,u) + A(u,t,s).
\end{eqnarray}
Unitarity may be monitored through the partial wave amplitudes $a_L^I$
for orbital angular momentum $L$,
\begin{equation}
a_L^I = {1\over 64\pi} \int_{-1}^{1}d({\rm cos}\theta)
P_L({\rm cos}\theta)T(I),
\end{equation}
with $T(I)=32\pi\Sigma (2L+1)P_L(\cos\theta)a_L^I$.
The unitarity condition $|2a_L^I-i|\le 1$ is sometimes approximated
by requiring $|a_L^I|\le 1$ or $|{\rm Re}\, a_L^I|\le{1\over 2}$.

Various models for these scattering amplitudes have been suggested
\cite{reviews,hantalk,bagger}.
We shall concentrate on models resulting from  effective chiral Lagrangians,
with and without  resonances, as follows.

\medskip\noindent
\underline{\bf (a) SM Heavy Higgs Model}

The Equivalence Theorem \cite{gbet}  gives the amplitude
\begin{equation}
A(s,t,u)= {-m_H^2 \over v^2}
\left(1+\frac{m_H^2}{s-m_H^2+im_H\Gamma_H\theta(s)}\right), \label{eq:A(stu)}
\end{equation}
where $m_H$ and $\Gamma_H$ are the Higgs boson mass and width,
$v=246$ GeV is the usual vacuum
expectation value, and $\theta(s)=1\ (0)$ for $s>0\ (s<0)$.
In all models, $A(t,s,u)$ and $A(u,t,s)$ are obtained by  permuting $s,t,u$.

\goodbreak
\medskip\noindent
\underline{\bf (b) Low-Energy Theorem (LET) Model \cite{let}}

This simply extrapolates the amplitudes, prescribed at low energy in
terms of   $v$
\begin{equation}
A(s,t,u) = s/v^2,
\end{equation}
and is the $m_H\to \infty$ limit of Eq.~(\ref{eq:A(stu)}).
This model eventually violates unitarity;
e.g.\ $a_0^0$ violates the bound  $|{\rm Re}\, a_0^0|\le{1\over 2}$
when $\sqrt s>1.2$ TeV and the less stringent bound $|a_0^0|\le 1$
when $\sqrt s>1.7$ TeV, where $s$ denotes the $WW$ invariant
mass squared.  Our present illustrations scarcely approach these non-unitary
ranges.  However, at higher energies we can unitarize the amplitudes
by a cut-off or by the $K$-matrix prescription
\begin{equation}
a_L^I \to a_L^I/(1-ia_L^I),
\end{equation}
which enforces the elastic unitarity condition $|2a_L^I -i|=1$.

\medskip\noindent
\underline{\bf (c) Chirally-Coupled Scalar (CCS) Model~\cite{bagger}}

This model describes the low-energy behaviour of a technicolor-type model
\cite{tclane} with a techni-sigma scalar resonance, through the amplitude
\begin{equation}
A(s,t,u)={s\over v^2}-({g_S^2s^2\over v^2}){1\over
s-M_S^2+iM_S\Gamma_S\theta(s)},
\end{equation}
where $M_S$ is the scalar resonance mass and
$\Gamma_S=3g_S^2M_S^3/(32\pi v^2)$ is its decay width into Goldstone fields.
The SM amplitude with $S=H$ is recovered for  $g_S^{} =1$.
We choose $M_S=1.0$ TeV and $\Gamma_S =0.35$ TeV, for which
$g_S^{}\simeq 0.84$; unsurprisingly, the results are similar to the
SM case.

\medskip\noindent
\underline{\bf (d) Chirally-Coupled Vector (CCV) Model~\cite{bagger,bess}}

This model describes the low-energy behaviour of a technicolor-type model
\cite{tclane} with a techni-rho vector resonance $V$, through the amplitude
\begin{equation}
A(s,t,u)={s\over 4v^2}(4-3a)+{aM_V^2\over 4v^2} \left[{u-s\over
t-M_V^2+iM_V\Gamma_V\theta(t)} + {t-s\over
u-M_V^2+iM_V\Gamma_V\theta(u)}\right],
\end{equation}
where $M_V$ and $\Gamma_V$ are the vector resonance mass and
width while $a=192\pi v^2\Gamma_V/M_V^3$.
We choose the case  $M_V=1.0$~TeV with $\Gamma_V =30$~GeV.
Note that the cross section for vector resonance production
increases as the width $\Gamma_V$ becomes larger.
The choice of a rather narrow width in our study is motivated from
LEP-I constraints via the $Z-V$ mixing \cite{besslimit}.

In our signal calculations, we will concentrate on the processes
of Eq.~(\ref{eq:WWfusion}),
which  go via the $W^+W^-$ initial state,
since the charged current coupling to the electron is larger than the
neutral current coupling.
We can calculate model (a)  directly from the complete SM
amplitudes without recourse to the Goldstone boson scattering
amplitude $A(s,t,u)$; the latter is shown above simply for comparison.
The same is true for model (b), in regions where unitarity is respected.
For these two cases, we define the SEWS signals  as the excesses
of heavy Higgs boson results over that of $m_H=0$.
Models (c) and (d) must however be calculated from the $A(s,t,u)$
expressions, using the Effective W-boson Approximation \cite{ewa}.

We emphasize that  the ratio of $W^+ W^- \to W^+ W^-$ and $W^+W^- \to ZZ$
cross sections is a sensitive probe of the SEWS \cite{hantalk}, since
the models have distinctive particle spectra with different
weak isospin content.  For a scalar-dominance model,
one expects the $W^+_L W^-_L$ rate to be larger than
$Z_LZ_L$;  {\it  e.g.}  a SM-like Higgs boson dominating
in the $s$-channel gives
$\sigma(H \rightarrow W^+_LW^-_L)/\sigma(H \rightarrow  Z_LZ_L) \sim$ 2.
For a vector-dominance model there would be a significant
resonant enhancement  in the $W^+_L W^-_L$ mode, but not in $Z_L Z_L$
due to the weak isospin conservation in  SEWS
(just like $\rho^0 \to \pi^+ \pi^-$ but not $\pi^0\pi^0$ in QCD).
On the other hand, if the resonances
are far from our reach, then the LET amplitudes behave like
$-u/v^2$ for $W_L^+W_L^-\to W_L^+W_L^-$
and like $s/v^2$ for $W_L^+W_L^-\to Z_LZ_L$, so that
$\sigma(W_L^+W_L^- \to Z_LZ_L)/\sigma(W_L^+W_L^- \to W_L^+W_L^-)=3/2$.
The $Z_L Z_L$ rate is then larger than $W^+_LW^-_L$,
and even more so in the central scattering region.
Measuring the relative yields of $W^+_L W^-_L$ and $Z_L Z_L$
will therefore reveal important characteristics of the SEWS.

\section{Dijet mass resolution}

We consider $W^\pm$ and $Z$ bosons detected by their dijet
decay modes and identified via the dijet invariant masses
$M(W^\pm \to jj)\simeq M_W$, $M(Z\to jj)\simeq M_Z$.  With realistic mass
resolution, discrimination cannot be made event-by-event but
can be achieved on a statistical basis.

   The experimental $W$ dijet mass distributions
will contain the intrinsic decay widths folded with experimental
resolution factors depending on calorimetry and geometry.
We have explored the possible dijet mass resolution using two alternative jet
energy resolution algorithms~\cite{jlc}
\begin{eqnarray}
\delta E_j/E_j &=& 0.50 \Big/ \sqrt{E_j} \;\oplus\; 0.02 \qquad
\mbox{Algorithm A} \\
               &=& 0.25 \Big/ \sqrt{E_j} \;\oplus\; 0.02 \qquad
\mbox{Algorithm B}
\end{eqnarray}
in GeV units, where the symbol $\oplus$ means adding in quadrature. We applied
this to the typical SM background process $e^+e^-\to e^+\nu W^-Z$ at $\sqrt s =
1.5$~TeV, averaging over all final $W\to jj$
dijet decays with
gaussian smearing of jet energies according to these algorithms; the resulting
$W^\pm \to jj$ and $Z\to jj$ dijet invariant mass distributions are shown in
Fig.~\ref{fig:dijetmass}. Since this study omits angular resolution effects,
sensitive to details of detector design, we shall adopt the more conservative
algorithm~A for further illustrations.

   If we now identify dijets having measured mass in the intervals
$$[0.85M_W, \; {1\over 2}(M_W+M_Z)] \quad
{\rm and}  \quad [{1\over 2}(M_W+M_Z), \; 1.15M_Z]$$
as $W^\pm \to jj$ and $Z\to jj$, respectively, algorithm A indicates that
true $W^+ W^-$, $W^\pm Z$, $ZZ\to jjjj$ events will be interpreted
statistically as follows:
$$
\begin{array}{lcrrrrr}
WW &\Rightarrow & 78\%\: WW, & 18\%\: WZ, &  1\%\: ZZ,&  3\%\: {\rm reject},
\\
WZ &\Rightarrow & 11\%\: WW, & 77\%\: WZ, & 9\%\: ZZ,&  3\%\: {\rm reject},
\\
ZZ &\Rightarrow &  2\%\: WW, & 22\%\: WZ, & 72\%\: ZZ,&  4\%\: {\rm reject},
\end{array}
$$
These numbers show that misidentification of $W^+ W^-$
as $ZZ$ (or vice versa)
is very unlikely; also the loss of $W^+ W^-$ or $ZZ$ signal strength is not
in itself very serious. The principal danger comes from $W^\pm Z$ events
that are misidentified as $W^+ W^-$ or $ZZ$, confusing or even swamping
these signals if $W^\pm Z$ production is relatively large.  We must therefore
ensure, via suitable acceptance criteria, that $W^\pm Z$ production is not
an order of magnitude bigger than $W^+ W^-$ or $ZZ$ signal.

A final caveat: the numbers above refer strictly to light-quark jets.
In $b$- and $c$-quark jets there is an appreciable probability of
$b\to c\ell\nu$  and/or  $c\to s\ell\nu$  ($\ell=e,\mu \hbox{ or }\tau$)
semileptonic decays, where
neutrinos deplete the visible jet energy.   Thus more $Z\to jj$ dijets
will be interpreted as $W^\pm \to jj$, but {\it not} vice versa.  We have
modeled this effect in typical situations with Scenario~A and find
that the correction to
the $W^\pm \to jj$ results is rather small.  However, about 8$\%$  more
$W^\pm Z$ events are now identified as $W^+ W^-$ (increasing this source of
background); also about 10$\%$ more $ZZ$ events are now
identified as $W^\pm Z$ (increasing this loss of signal).
These changes are significant but not disastrous.
The resulting modified identification probabilities are as follows:
$$
\begin{array}{lcrrrrr}
WW &\Rightarrow & 73\%\: WW, & 17\%\: WZ, &  1\%\: ZZ,&  9\%\: {\rm reject},
\\
WZ &\Rightarrow & 19\%\: WW, & 66\%\: WZ, & 7\%\: ZZ,&  8\%\: {\rm reject},
\\
ZZ &\Rightarrow &  5\%\: WW, & 32\%\: WZ, & 55\%\: ZZ,&  8\%\: {\rm reject},
\end{array}
$$
and we will use these numbers  in the rest of  our analyses.

When the dijet mass resolution function is known, for a given detector,
the apparent $W^+ W^-,W^\pm Z$ and  $ZZ$ rates can be unfolded to determine
approximately
the underlying true rates.  In the following, we first concentrate our
attention on these true rates,  and then use the examples above to
estimate resolution effects.

\section{SM calculations and acceptance cuts}

The SM signals for  $W_L^+ W_L^-\to W_L^+W_L^-,\; Z_LZ_L$ fusion processes
with a heavy Higgs boson have been considered previously, along with certain
SM backgrounds~\cite{fusionjfg,hagi,fusionvb,najima}.
The irreducible SM backgrounds to the Strongly-interacting
Electro-Weak Sector,  which include transversely polarized
vector bosons  $W_T^\pm$ and  $Z_T$ production,
can be obtained by setting $m_H = 0$; further backgrounds
arise from misidentifying other SM channels in Eq.~(2).

We first address the annihilation background of Eq.~(3), which has different
dynamics and requires different cuts from the other backgrounds.
It is important at $\sqrt{s}=0.5$ TeV~\cite{fusionvb}, but unlike the
scattering
channels, its cross section increases slowly with $\sqrt{s}$ above
the threshold and then decreases after $\sqrt{s} \sim 1$ TeV,
like $1/s$ asymptotically.
At $\sqrt{s}=1.5$ TeV the total annihilation background cross section
is of order 10 fb (taking into account three flavors of the neutrinos
from the $Z$ decay),  comparable to the SEWS signals.
It can be reduced severely, however, by a cut on the recoil mass
$M_{\rm recoil}$ that is the invariant mass of all the final-state particles
excluding the $W^+W^-\to (jj)(jj)$ system:
\begin{equation}
M_{\rm recoil}^2 = s + M_{WW}^2 - 2 \sqrt{s} (E_{W^+} + E_{W^-} ) \;,
\label{eq:recoilmass}
\end{equation}
%
where the $W$ boson energies $E_W$ are defined in the $e^+e^-$ CM frame
and $\sqrt{s}$ is the CM energy of the $e^+e^-$ collider.
The recoil-mass spectrum of the annihilation Eq.~(3) peaks at $M_Z$,
due to the $Z \to \nu \bar \nu$ decay, but this peak is smeared out
by the contributions of initial-state radiation as well as the mismeasurement
of the $W$ hadronic energies.  A cut such as
\begin{equation}
M_{\rm recoil} > 200\; \rm GeV\; ,
\label{eq:rcut}
\end{equation}
therefore effectively suppresses the annihilation background.
Figure~\ref{fig:recoil} shows that this background is reduced
to about 1--2\% of typical SEWS signals, with negligible effect on the
scattering channels at $\sqrt s=1.5$ TeV.  We shall henceforth make this
cut and neglect the annihilation background of Eq.~(3).

The remaining scattering cross sections of interest are illustrated
in Fig.~\ref{fig:sews}.  This figure shows SM cross sections for the fusion
processes with both $m_H=0$ (solid curves) and $m_H=1$ TeV (dashed
curves);  the excess over the $m_H=0$
case represents the SEWS signal in the SM Heavy Higgs Model.
The SEWS signals of present interest have final-state $W_L^+ W_L^-$ and
$Z_LZ_L$ pairs, giving four-jet final states with two undetected
neutrinos [see Eq.~(\ref{eq:WWfusion})]; the branching fractions for
four-jet decays  are not included in this  section.

We start with the most basic acceptance cuts.  Since we are interested
in $WW$ scattering at high subprocess energy, we look for
pairs of weak bosons with high invariant masses $M_{WW}$, high transverse
momenta $p_T(W)$ of the vector bosons,
and relatively large angles $\theta_W$ with respect to the beam axis.  We
require
\begin{equation}
M_{WW} >500{\rm\ GeV} \;;
\quad p_T(W) > 150{\rm\ GeV} \;;
\quad |\cos\theta_W| <0.8.
\label{eq:level1}
\end{equation}
The solid curves in Fig.~\ref{fig:mwwmzz} show the resulting cross
sections at $\sqrt s=1.5$ TeV in (a) the $W^+W^-$ channel and (b) the $ZZ$
channel,  for the SM Heavy Higgs Model ($m_H=1$ TeV) and the LET Model
($m_H=\infty$); SM backgrounds are also shown.   Note that the solid
curves represent sums of signal plus the intrinsic background; the signals
alone are found by subtracting the $m_H=0$ curve, in this and
subsequent figures.

Figure~\ref{fig:mwwmzz} immediately illustrates
the main point of this paper, that the $W^+W^-/ZZ$ signal ratio is
sensitive to the details of SEWS; we see that the SM Heavy Higgs
model ($m_H=1$ TeV) gives $W^+W^-/ZZ > 1$ whereas the LET Model ($m_H=\infty$)
gives $W^+W^-/ZZ < 1$.  It also shows that the $ZZ$ signals around $M_{ZZ}
\sim 1$ TeV are bigger than the backgrounds, since
Section III indicates that the $W^+W^-$
background has very small ($\sim 1 \%$) probability to be misidentified
as $ZZ$, but more work is needed to separate the $W^+W^-$ signals.

The SM $e^+e^-W^+W^-$ background gets very large contributions from
the virtual $\gamma\gamma\to W^+W^-$ subprocess, which gives
mainly dibosons with small net transverse momentum $p_T(WW)$, quite
unlike the SEWS signal and other backgrounds. Figure~\ref{fig:pT(WW)}
compares the $W^+W^-$ signals and backgrounds versus $p_T(WW)$, after
the first-level cuts of Eqs.~(\ref{eq:rcut})--(\ref{eq:level1});
it shows the small-$p_T$ peak of $e^+e^-W^+W^-$,
and also shows how the $W_T$ backgrounds
are favored at very large $p_T$.  It is clearly advantageous to select
an intermediate range of $p_T(WW)$, to remove a lot of background
at little cost to the signal;  we make somewhat similar cuts
for $p_T(ZZ)$ , though these are less crucial.  Specifically we require
\begin{equation}
50{\rm\ GeV} < p_T(WW) < 300{\rm\ GeV}, \; \; \;
20{\rm\ GeV} < p_T(ZZ) < 300{\rm\ GeV},
\label{eq:level2}
\end{equation}
at $\sqrt s =1.5$ TeV.
With large minimum $p_T(WW)$ and $p_T(ZZ)$ requirements, it becomes
much less likely that the final-state electrons in $eeWW$ and $e\nu WZ$
background channels can escape undetected down the beam-pipes; a veto
on visible hard electrons is now very effective against $eeWW$ (less so
against $e\nu WZ$).  We therefore impose the veto \cite{hagi}
\begin{equation}
\mbox{no $e^\pm$ with $E_e>50$ GeV and} \quad
|\cos\theta_e| < \cos(0.15\rm\ rad)
\label{eq:level3}
\end{equation}

Figure~\ref{fig:mH=1} compares the resulting $m_H=1$ TeV (SM) and
$m_H=\infty$ (LET) cross sections with backgrounds at $\sqrt s=1.5$ TeV,
versus diboson invariant mass, after imposing all the above cuts.
Combining these results with the typical $WW \Rightarrow ZZ$ and
$WZ \Rightarrow WW,ZZ$ misidentification probabilities from
Section III, we see that both SEWS model signals are now
observable  over
the total remaining SM backgrounds.  We henceforth adopt
the cuts of Eqs.~(\ref{eq:rcut})--(\ref{eq:level3})
and present detailed results in the next section.

The lowest order backgrounds $e^+e^-\to
W^+W^-,ZZ$ can be removed by the cuts on $p_T(WW)$ and $M_{\rm recoil}$.
We have  neglected
QCD backgrounds from $e^+e^-\to jjjj$ production.
They are formally of order $\alpha^2\alpha_s^2$ compared to our
electroweak cross sections of order $\alpha^4$ , but the QCD 4-jet
final states contain no direct neutrino production and will be
heavily suppressed by the $M_{\rm recoil}$ and $p_T(WW)$ cuts; they will
be further suppressed by the $M(jj)\simeq M_W,M_Z$ requirements.
We have also neglected $e^+e^-\to \bar tt \to \bar bbW^+W^-$
as a source of background; this gives unwanted extra jets and
would be suppressed by the $p_T(WW)$ cut if the b-jets escaped
near the beam axis.

\section{Results}

\begin{table}
\def\arraystretch{.8} 
\caption[]{\label{tablei}\small
Cross sections in fb, before and after cuts, for $e^+e^-$ collisions
at $\protect\sqrt s=1.5$ TeV.
For comparison, results for $e^-e^- \rightarrow \nu \nu W^-W^-$
are also presented, with the same energy and the $W^+W^-$ cuts.
Hadronic branching fractions of $WW$ decays and the
$W^\pm/Z$ identification/misidentification  are not included here.
The first number in the final $e^+e^-W^+W^-$ and $e\nu WZ$ entries
denotes the $p_T > 20$~GeV choice, for the case where $WW$ and $WZ$
are misidentified as $ZZ$; the second number (in parentheses) denotes
the $p_T > 50$~GeV choice, for the case where they are identified as $WW$.}
\[
\begin{array}{|l|c|c|c|}
\hline
\mbox{Contribution}&\mbox{no cuts}&
\mbox{with Eqs.~(\ref{eq:rcut})--(\ref{eq:level1})}&
\mbox{with Eqs.~(\ref{eq:rcut})--(\ref{eq:level3})}\\
\hline
\bar\nu\nu W^+W^-\mbox{ signals (fb)} & & & \\
\quad\mbox{SM }(m_H^{}=1\mbox{ TeV})& 7.7   & 3.5      & 2.4 \\
\quad\mbox{CCS }(M_S^{},\Gamma_S=1,0.35 \mbox{ TeV})  & - & 3.5  &2.4  \\
\quad\mbox{CCV }(M_V^{},\Gamma_V=1,0.03 \mbox{ TeV})  & - & 1.5   & 1.0  \\
\quad\mbox{LET }(m_H^{}=\infty)&3.1 & 0.61         & 0.46              \\
\hline
\bar\nu\nu ZZ\mbox{ signals (fb)} & & & \\
\quad\mbox{SM }(m_H^{}=1\mbox{ TeV})& 5.9   & 2.4  & 2.2 \\
\quad\mbox{CCS }(M_S^{},\Gamma_S=1,0.35 \mbox{ TeV})  & - & 2.7  &   2.5 \\
\quad\mbox{CCV }(M_V^{},\Gamma_V=1,0.03 \mbox{ TeV})  & - &  0.72  & 0.67 \\
\quad\mbox{LET }(m_H^{}=\infty)&3.4 & 0.89          & 0.84            \\
\hline
\nu\nu W^-W^-\mbox{ signals (fb)} & & & \\
\quad\mbox{SM }(m_H^{}=1\mbox{ TeV})& 2.7  & 0.53      & 0.39 \\
\quad\mbox{CCS }(M_S^{},\Gamma_S=1,0.35 \mbox{ TeV})  & - & 0.71    & 0.52   \\
\quad\mbox{CCV }(M_V^{},\Gamma_V=1,0.03 \mbox{ TeV})  & - & 0.72 & 0.53 \\
\quad\mbox{LET }(m_H^{}=\infty)& 3.5 &   0.89 & 0.63  \\
\hline
\mbox{SM Backgrounds (fb)} & & & \\
\quad\bar\nu\nu W^+W^-\;(m_H^{}=0)    & 45 & 1.1 & 0.86 \\
\quad\bar\nu\nu ZZ\;(m_H^{}=0)    & 18  & 0.84 & 0.72 \\
\quad e^+e^-W^+W^-\;(m_H^{}=0)& 2000  & 28 & 3.5 (0.95) \\
\quad e\nu WZ\;(m_H^{}=0)     &  150  & 4.6  & 3.1 (2.7) \\
\quad e^-e^- \rightarrow \nu \nu W^-W^-  \;
(m_H^{}=0)& 51  & 2.3 & 1.7 \\
\hline
\end{array}
\]
\end{table}

Table~\ref{tablei} presents our results  for $e^+e^-$ collisions at $\sqrt s =
1.5$ TeV, showing signal and  background cross sections before and
after successive cuts.   Here the SM Heavy Higgs and LET Model
signals have been found by subtracting the SM $m_H=0$ intrinsic background
from SM $m_H=1$ TeV and $m_H=\infty$ values, respectively. Partial wave
unitarity is respected at all energies reached so that no unitarization needs
to be imposed~\cite{peak}.
For the chirally coupled scalar (CCS) and chirally coupled
vector (CCV) models, the signals are calculated in the Effective $W$-boson
Approximation \cite{ewa}. The validity of this approximation can be checked by
comparing CCS (with $g_S=1$) to the  exact SM  results;
there is agreement at the $20\%$ level,  using the
cuts in Eq.~(\ref{eq:level1}).  In such an approximation, however, the
kinematical cuts
of Eqs.~(\ref{eq:level2})--(\ref{eq:level3}) cannot be implemented;  we have
therefore assumed
the efficiencies of these cuts to be the same as for the SM heavy Higgs
boson ($m_H=1$ TeV) signal.
For comparison, results for  $e^-e^- \rightarrow \nu\nu W^-W^-$
are also included \cite{bbch}, with the same cuts as the $\bar\nu\nu W^+W^-$
case.   We remark that the LET signal rates for  $e^+e^-\to \nu\bar\nu ZZ$ and
$e^-e^-\to \nu\nu W^-W^-$ channels are essentially equal
(when the cuts imposed are
the same); this is a consequence of  the Low Energy Theorem and crossing
symmetry for $W_LW_L$ scattering.
Branching fractions  for $W \to jj$ decays and
$W^\pm/Z$  identification/misidentification factors
are not included in this  table.

In Fig.~\ref{fig:events} we present the expected signal
and background event rates versus diboson mass for different models at
a 1.5~TeV NLC, assuming an integrated luminosity of 200 fb$^{-1}$.
The branching fractions  $BR(W \to jj)=67.8\%$ and $BR(Z\to jj)=69.9\%$
\cite{pdg} and the
$W^\pm /Z$  identification/misidentification factors (final set of
Section~III) are all included here.
Comparing the $W^+ W^-$ events (Fig.~7a) and $ZZ$ events (Fig.~7b),
we once again see that a broad Higgs-like scalar will enhance both
$W^+ W^-$
and $ZZ$ channels with $\sigma(W^+ W^-) > \sigma(ZZ)$; a $\rho$-like vector
resonance will manifest itself through $W^+W^-$ but not $ZZ$; while the LET
amplitude will enhance $ZZ$ more than $W^+ W^-$.
Table~\ref{tableii} summarizes the corresponding total signal $S$ and
background $B$
event numbers, summing over diboson invariant mass bins,  together with
the statistical significance $S/\sqrt B$.  The  LET signal for  $W^+W^-$  is
particularly small; the ratio $S/B$ can be enhanced by making a higher
mass cut (e.g.\ $M_{WW} > 0.7$ TeV), but the significance $S/\sqrt B$
is not in fact improved by this.
Results for  $e^-e^- \rightarrow \nu\nu W^-W^-$ have again been
included for comparison.

At the NLC, since electron polarization of order 90--95\% at injection
with only a few percent depolarization during acceleration may
well be achievable \cite{nlcetc},
it is interesting to consider also the effects of beam polarization.
The $W^+W^- \to W^+ W^-,ZZ$ scattering signals of interest arise
from initial $e^-_L$ and $e^+_R$ states only and the signal cross sections
are therefore doubled with an $e^-_L$ beam.
Table~\ref{tableiii}(a)  shows the background
cross sections for the beams $e^+ e^-_L$, $e^- e^-_L$
and $e^-_L e^-_L$.  Based on these results, event numbers and significances
for the case of 100\% $e^-_L$ beam at  $\sqrt s=1.5$ TeV with 200 fb$^{-1}$
are shown in Table~\ref{tableiii}(b), to be compared with Table~\ref{tableii};
$S$ and $B$ for intermediate beam polarizations
can be found by interpolating Table~\ref{tableii} and Table~\ref{tableiii}.

\begin{table}[h]
\centering
\caption[]{\label{tableii}\small
Total numbers of $W^+W^-, ZZ \rightarrow  4$-jet
signal $S$ and background $B$ events calculated for  a 1.5~TeV
NLC with  integrated luminosity 200~fb$^{-1}$.  Events are summed
over the mass range $0.5 < M_{WW} < 1.5$~TeV except for the $W^+W^-$ channel
with  a narrow vector resonance in which $0.9 < M_{WW} < 1.1$~TeV. The
statistical significance $S/\sqrt B$ is also given.
For comparison, results for $e^-e^- \rightarrow \nu \nu W^-W^-$
are also presented, for the same energy and luminosity and the $W^+W^-$
cuts. The hadronic branching fractions of $WW$ decays and the $W^\pm/Z$
identification/misidentification are included.}
\medskip
\begin{tabular}{|l|c|c|c|c|c|} \hline
channels & SM  & Scalar & Vector   & LET  \\
\noalign{\vskip-1ex}
& $m_H=1$ TeV & $M_S=1$ TeV & $M_V=1$ TeV &\\
\hline
$S(e^+ e^- \to \bar \nu \nu W^+ W^-)$
& 160   & 160   & 46  & 31  \\
$B$(backgrounds)
& 170    & 170   & 4.5  & 170  \\
$S/\sqrt B$ & 12 & 12 & 22 & 2.4 \\
\hline
$S(e^+ e^- \to \bar\nu \nu ZZ)$
&  120  & 130  & 36  & 45   \\
$B$(backgrounds)
& 63    & 63   & 63  & 63  \\
$S/\sqrt B$ & 15& 17& 4.5& 5.7\\
\hline
\hline
$S(e^- e^- \to \nu \nu W^- W^-)$
& 27  & 35  & 36  & 42  \\
$B$(backgrounds)
& 230  & 230   & 230  & 230  \\
$S/\sqrt B$ & 1.8 & 2.3 & 2.4 & 2.8 \\
\hline
\end{tabular}
\end{table}

\begin{table}
\def\arraystretch{.8}
\centering
\smallskip
\caption[]{\label{tableiii} \small
Improvements from using $100\%$ polarized $e^-_L$ beams in a 1.5~TeV
$e^+e^-/e^-e^-$ collider.
Part (a) gives SM background cross sections in fb with the full cuts
Eqs.~(\ref{eq:rcut})--(\ref{eq:level3}); the signal cross sections are
simply doubled with each $e^-_L$ beam  compared to Table~\ref{tablei}.
Part (b) gives the expected numbers of signal and background events
for integrated luminosity 200~fb$^{-1}$ , to be compared with
Table~\ref{tableii}.
}
\begin{tabular}{|l|c|}
\hline
(a) SM Backgrounds & Cross sections in fb with
Eqs.~(\ref{eq:rcut})--(\ref{eq:level3}) \\
\hline
$\quad e^+e^-_L \to \bar \nu \nu W^+ W^- \; (m_H=0)$ & 1.7 \\
$\quad e^+e^-_L \to \bar \nu\nu ZZ \; (m_H=0)$   &    1.4 \\
$\quad e^+e^-_L \to e^+e^- W^+W^-\; (m_H=0)$ &         4.3 (1.3) \\
$\quad e^+e^-_L \to e\nu WZ\; (m_H=0)$ &               4.5 (3.9) \\
\hline
\hline
$\quad e^-e^-_L \to \nu \nu W^- W^- \; (m_H=0)$ &  3.4 \\
$\quad e^-e^-_L \to e^-e^- W^+ W^- \; (m_H=0)$ &  1.3 \\
$\quad e^-e^-_L \to e^-\nu  W^- Z \; (m_H=0)$ &   4.4 \\
\hline
$\quad e^-_L e^-_L \to \nu \nu W^- W^- \; (m_H=0)$ &  6.8 \\
$\quad e^-_L e^-_L \to e^-e^- W^+ W^- \; (m_H=0)$ &  1.8 \\
$\quad e^-_L e^-_L \to e^-\nu  W^- Z \; (m_H=0)$ &   6.5 \\
\hline
\end{tabular}

\vspace{0.05in}

\begin{tabular}{|l|c|c|c|c|}
\hline
(b) channels & SM  & Scalar & Vector   & LET  \\
& $m_H=1$ TeV & $M_S=1$ TeV & $M_V=1$ TeV &\\
\hline
$S(e^+ e^- \to \bar \nu \nu W^+ W^-)$
& 330   & 320   & 92  & 62  \\
$B$(backgrounds)
& 280    & 280   & 7.1  & 280  \\
$S/\sqrt B$ & 20 & 20 & 35 & 3.7 \\
\hline
$S(e^+ e^- \to \bar\nu \nu ZZ)$
&  240  & 260  & 72  & 90   \\
$B$(backgrounds)
& 110    & 110   & 110  & 110  \\
$S/\sqrt B$ & 23 & 25& 6.8& 8.5\\
\hline
\hline
$S(e^- e^-_L \to \nu \nu W^- W^-)$  & 54 & 70 & 72 & 84 \\
$B$(background) & 400 & 400 & 400 & 400\\
$S/\sqrt B$ & 2.7 & 3.5 & 3.6 & 4.2 \\
\hline
$S(e^-_L e^-_L \to \nu \nu W^- W^-)$  & 110 & 140 & 140 & 170 \\
$B$(background) & 710 & 710 & 710 & 710\\
$S/\sqrt B$ & 4.0 & 5.2 & 5.4 & 6.3 \\
\hline
\end{tabular}
\end{table}

Since the SEWS signals increase with  CM energy,
a  2 TeV $e^+e^-$ linear collider  would give a larger signal rate.
We find (see Fig.~\ref{fig:sews} )
that at $\sqrt s=2$ TeV the $m_H=1$ TeV and $m_H=\infty$
signal cross sections are, for $W^+W^-$,
\begin{eqnarray}
\sigma^{}_{\rm SEWS}({\rm SM}\; 1 {\rm TeV}) &=& \sigma_{W^+W^-}
(m_H=1 \; {\rm TeV})
- \sigma_{W^+W^-}(m_H=0) \simeq 20 \; {\rm fb} \nonumber \\
\sigma^{}_{\rm SEWS}({\rm LET}) &=& \sigma_{W^+W^-}(m_H=  {\infty})
- \sigma_{W^+W^-}(m_H=0) \simeq  5 \; {\rm fb}  \nonumber
\end{eqnarray}
and  for $ZZ$,
\begin{eqnarray}
\sigma^{}_{\rm SEWS}({\rm SM}\; 1 {\rm TeV}) &=& \sigma_{ZZ}(m_H=1 \; {\rm
TeV})
- \sigma_{ZZ}(m_H=0)  \simeq 14\; {\rm fb}  \nonumber \\
\sigma^{}_{\rm SEWS}({\rm LET}) &=& \sigma_{ZZ}(m_H=  {\infty})
- \sigma_{ZZ}(m_H=0)  \simeq7 \; {\rm fb} \nonumber
\end{eqnarray}
The signal rates are enhanced by about a factor $\sim$ 2--2.5
by increasing the CM energy from 1.5 to 2 TeV (compared with the
first numerical column in Table~\ref{tablei}).

It may be more
advantageous to study the SEWS at  possible higher energy $\mu^+\mu^-$
colliders~\cite{sausalito}.  To demonstrate this point,  Fig.~\ref{fig:mumu}
gives the $\sqrt s$-dependence of the corresponding SM total cross sections
for $m_H=1$ TeV as well as  the various
backgrounds (with $m_H=0$).  The excesses over the $m_H=0$ case again
represent the SEWS signals in the SM Heavy Higgs Model.
We see that the uncut $\bar\nu\nu W^+W^-$ and $\bar\nu\nu ZZ$ signals increase
most rapidly at the lower energies; starting from the NLC values
at $\sqrt s=1.5$ TeV, they have increased by factors $\sim{}$2--2.5
at $\sqrt s=2$ TeV  and by factors $\sim 10$
at $\sqrt s=4$ TeV (the value currently being discussed for a possible
$\mu^+\mu^-$ circular collider~\cite{sausalito}).
More detailed considerations would depend on
design parameters of the collider and detector; in the absence of firm
information, we do not pursue this question any further here~\cite{mumu}.

Finally we note that our calculated cross sections and event rates
neglect bremsstrahlung and beamstrahlung initial state radiation,
which somewhat reduce the effective CM energy and with it the signal
and principal backgrounds.  Colliders are usually designed to
minimize beamstrahlung. The net corrections are expected to be
small and  our general conclusions are not affected.

\section{Summary and Discussion}

\indent

Our main results are summarized in Fig.~\ref{fig:events} and
Tables~\ref{tableii}--\ref{tableiii}
for an $e^+e^-$ collider at $\sqrt s=1.5$~TeV.
They show that the $W^+W^-/ZZ$ event ratio is a sensitive probe of
SEWS dynamics.
Indeed, the differences between the various models are quite
marked and the observation of such signals would provide strong indications
about the underlying dynamics of the SEWS.
In fact, not only the ratio but also the size of the separate $W^+W^-$ and
$ZZ$ signals contains valuable dynamical information.
Our results show  statistically significant signals for a 1 TeV scalar or
vector state.  We also find a 5.7$\sigma$ signal for the LET amplitudes
via the $W^+W^- \to ZZ$ channel alone without the improvement
by beam polarization.  Our event numbers
are based on optimized acceptance cuts and a luminosity 200~fb$^{-1}$,
roughly corresponding to one year running with a favorable
design \cite{nlcetc,barklow}.

Our approach is based on $W^+W^-,ZZ\to (jj)(jj)$ four-jet signals, and
therefore relies on good dijet mass resolution.  Our simulations included
energy resolution but not angular resolution, being conservative about
the former to compensate for our neglect of the latter; we also
folded in the effects of finite $W$ and $Z$ widths and of
semileptonic decays in $b$- and $c$-quark jets.
We therefore believe that our final $W^\pm/Z$
identification/misidentification factors are not unrealistic.

For an $e^-e^-$ collider with the same energy and luminosity,
the LET signal rate  for the $\nu\nu W^-W^-$ ($I=2$) channel
is similar to the LET result of  $e^+e^-\to\bar\nu\nu ZZ$, as anticipated,
while the background rate is higher.

The signals are doubled for an $e^-_L$ polarized beam (or
quadrupled for two $e^-_L$ beams), whereas the backgrounds increase
by smaller factors.  Hence polarization improves the significance
of signals substantially, for given luminosity: compare
Tables II and III.

The signals also increase strongly with the CM energy.
A 2 TeV $e^+e^-$ linear collider would increase the signal rates
by roughly a factor of 2--2.5. If future $\mu^+\mu^-$
colliders can reach higher energies with comparable luminosities and
comparable signal/background discrimination, an order of magnitude
increase in signal rate may be expected at $\sqrt s=4$ TeV: see Fig.~8.

By way of further discussion, we offer the following comments and
comparisons.\\
{\bf (a)} It may be possible to exploit $ZZ\to (jj)(\ell^+\ell^-)$
signals, to confirm the hadronic $ZZ$ results.  The former have
smaller branching fraction (reducing both signal and intrinsic
background by a factor 0.19), but the $WZ\Rightarrow ZZ$ misidentification
background is reduced by a factor 0.1 and the $W^+W^-$ backgrounds
are eliminated.
 It may also be possible to exploit $b$-tagging to improve the
discrimination between $W^\pm$ and $Z$ dijets.  Since $39\%$ of all
$ZZ\to (jj)(jj)$ events contain at least one $b\bar b$ jet pair,
and $b$-tagging efficiencies of 30--40\% per event can be
contemplated, requiring a tag would reduce the $ZZ\to 4j$ signal
and intrinsic backgrounds by a factor 0.12--0.16; in comparison, $WZ$ events
would be reduced by 0.066--0.088 and $W^+W^-$ backgrounds
would be eliminated.\\
{\bf (b)} The direct $s$-channel process $e^+e^- \rightarrow W^+W^-$
should be more advantageous in searching for effects from a vector
$V$ through $\gamma,Z-V$ mixing \cite{bess,besslimit,barklow,peskin}, due to
more efficient use of the CM energy, the known beam energy constraint,
and better control of backgrounds. However, the $WW$ fusion
processes studied here involve more  spin-isospin channels
of $WW$ scattering;  they are unique
for exploring scalar resonances and are complementary to the direct
$s$-channel for the vector and non-resonant cases.\\
{\bf  (c)} The conclusions of Ref.~\cite{kurihara} are pessimistic about
studying the LET amplitude ($m_H \rightarrow \infty$) at
a 1.5 TeV NLC via the $\bar\nu \nu W^+W^-$ channel; in contrast,
we find that the NLC has significant potential to explore
non-resonant SEWS physics, reaching about a 5.7$\sigma$ signal
for one-year running in the $\bar\nu \nu ZZ$ channel alone.
The improvement comes mainly
from including the $W^+ W^- \to  ZZ$ process and from our optimized
kinematical cuts to suppress the backgrounds while maximally preserving
the SEWS signal. \\
{\bf  (d)} We have concentrated on $\nu\bar \nu W^+W^-, \nu\bar \nu ZZ$ final
states,
neglecting $e\nu WZ$ signals, because the heavy Higgs
boson  contribution to the
latter is negligibly small compared to the irreducible SM background
(see Fig.~\ref{fig:sews}). However, the presence of an $I=1$ vector
state would greatly enhance the cross section for
$e^+e^- \rightarrow e^\pm\nu W^\mp Z$ \cite{kurihara}. Our optimal
kinematical cuts and the $M(jj)$ reconstruction should be essentially
applicable to the $WZ$ channel and a wider study including this channel
would provide consistency checks on SEWS effects.\\
{\bf (e)} The LET amplitudes we employed correspond to the lowest order
universal term in the energy expansion in effective chiral
Lagrangians \cite{chirall}. The magnitude and sign of coefficients
of higher dimension operators (the so-called anomalous couplings)
in the Lagrangians would depend on specific SEWS models.
In the clean environment at the NLC, one may be
able to measure the shape as well as the normalization of the
$WW$ mass distribution rather well. If this can be achieved,
one may even hope to study the non-resonance amplitudes in
detail to go beyond the LET term and
to extract the underlying dynamics at
higher mass scales beyond $\cal O$(1 TeV). \\
{\bf (f)} Finally, in studying a scalar or a vector resonance, we have
followed the simplest approach of assuming just one resonance
at a time. It has been emphasized recently \cite{hhh} that
there may coexist several resonances (as in low energy QCD),
a scalar ($\sigma$-like),  a vector ($\rho$-like), an axial vector
($a_1$-like) and an isospin-singlet vector ($\omega$-like),
obeying some algebraic relations to satisfy the proper Regge behavior
and certain sum rules of strong scattering \cite{hadrons}.
There would be definite relations among the masses and couplings
of these resonances, leading to cancellation and other predictions
in the  strong scattering amplitudes. This possibility deserves
further scrutiny in studying SEWS effects at colliders.

\section*{Acknowledgments}
We thank Tim Barklow, Dave Burke and Ron Ruth for helpful conversations
on many NLC issues. We are grateful to  Jonathan Feng,
Jon Guy and Bill Scott for discussions about dijet mass resolution.
This work has been supported in part by Department of Energy
Grants DE-FG03-91ER40674, DOE-FG03-93ER40757 and
DE-FG02-95ER40896,  and in part
by the University of Wisconsin Research Committee with funds granted by the
Wisconsin Alumni Research Foundation.
T.H. is also supported in part by a UC-Davis Faculty Research Grant.

\newpage

\section*{Figure captions}
\renewcommand{\labelenumi}{Fig.~\arabic{enumi}:}

\begin{enumerate}

\item{\label{fig:dijetmass}
$W^\pm \to jj$ and $Z\to jj$ dijet invariant mass distributions for
$e^+e^- \to e\nu WZ$ events at $\sqrt s=1.5$ TeV, found by applying
(a)~algorithm~A and (b)~algorithm~B (see text) for calorimeter energy
resolution, omitting angular resolution and heavy-quark decay  effects.}
\item{\label{fig:recoil}
SM $e^+e^-\to \bar\nu\nu W^+ W^-$ annihilation and scattering
cross sections versus $M_{\rm recoil}$ at $\sqrt s=1.5$ TeV.
Annihilation (dotted curve) is compared to scattering with $m_H=1$ TeV
(solid curve) and $m_H=0$ (dashed curve).}
\item{\label{fig:sews}
Cross sections for SM scattering processes that can contribute SEWS
signals and backgrounds in the $e^+e^-\to \bar\nu\nu W^+W^-$ and $\bar\nu\nu
ZZ$ channels, versus CM energy $\sqrt s$.}

\item{\label{fig:mwwmzz}
SEWS signal and background cross sections  versus diboson invariant
mass at $\sqrt s=1.5$ TeV, after the first-level cuts of
Eqs.~(\ref{eq:rcut})--(\ref{eq:level1}), in the channels
(a) $e^+e^- \to \bar\nu\nu W^+W^-$
and (b) $e^+e^- \to \bar\nu\nu ZZ$.   Solid curves denote total SM
contributions with $m_H=1$ TeV (Heavy Higgs Model) and with $m_H=\infty$
(LET Model); dotted curves denote intrinsic SM backgrounds ($m_H=0$).
Dashed and dot-dashed curves show $eeWW$ and $e\nu WZ$ production.
$W^\pm ,Z\to jj$ branching fractions and $W^\pm/Z$
identification/misidentification factors are not included.}

\item{\label{fig:pT(WW)}
$W^+W^-$ signal and background cross sections versus transverse
momentum $p_T(WW)$ after the first-level cuts of
Eqs.~(\ref{eq:rcut})--(\ref{eq:level1}).
Solid curves denote total contributions
from the SM Heavy Higgs Model (with $m_H=1$ TeV) and the LET Model
(with $m_H=\infty$); other curves denote backgrounds as in Fig.~4.
$W^\pm ,Z\to jj$ branching fractions and $W^\pm/Z$
identification/misidentification factors are not included.}

\newpage

\item{\label{fig:mH=1}
SEWS signal and background cross sections versus diboson
invariant mass at $\sqrt s=1.5$~TeV, after the combined cuts
of Eqs.~(\ref{eq:rcut})--(\ref{eq:level3}):
(a) in the $W^+W^-$ channel and (b) in the $ZZ$ channel.  Notation
follows Fig.~4.  $W^\pm,Z\to jj$ branching fractions and $W^\pm/Z$
identification/misidentification factors are not included.}
\item{\label{fig:events}
Expected numbers of $W^+W^-,ZZ\to (jj)(jj)$ signal and background
events, in 20 GeV bins of diboson invariant mass, for 200~fb$^{-1}$
luminosity at $\sqrt s=1.5$ TeV: (a) $W^+W^-$ events, (b) $ZZ$ events.
Dijet branching fractions and $W^\pm/Z$ identification/misidentification
factors are included.  The dotted histogram denotes total SM
background including misidentifications.  The solid, dashed and dot-dashed
histograms denote signal plus background for the LET, SM and CCV models,
respectively; CCS model results are close to the SM~case.}
\item{\label{fig:mumu}
Cross sections for SM scattering processes that contribute SEWS signals and
backgrounds in the $\mu^+\mu^-\to \bar\nu\nu W^+W^-$ and $\bar \nu\nu ZZ$
channels, versus CM energy $\sqrt s$.}

\end{enumerate}

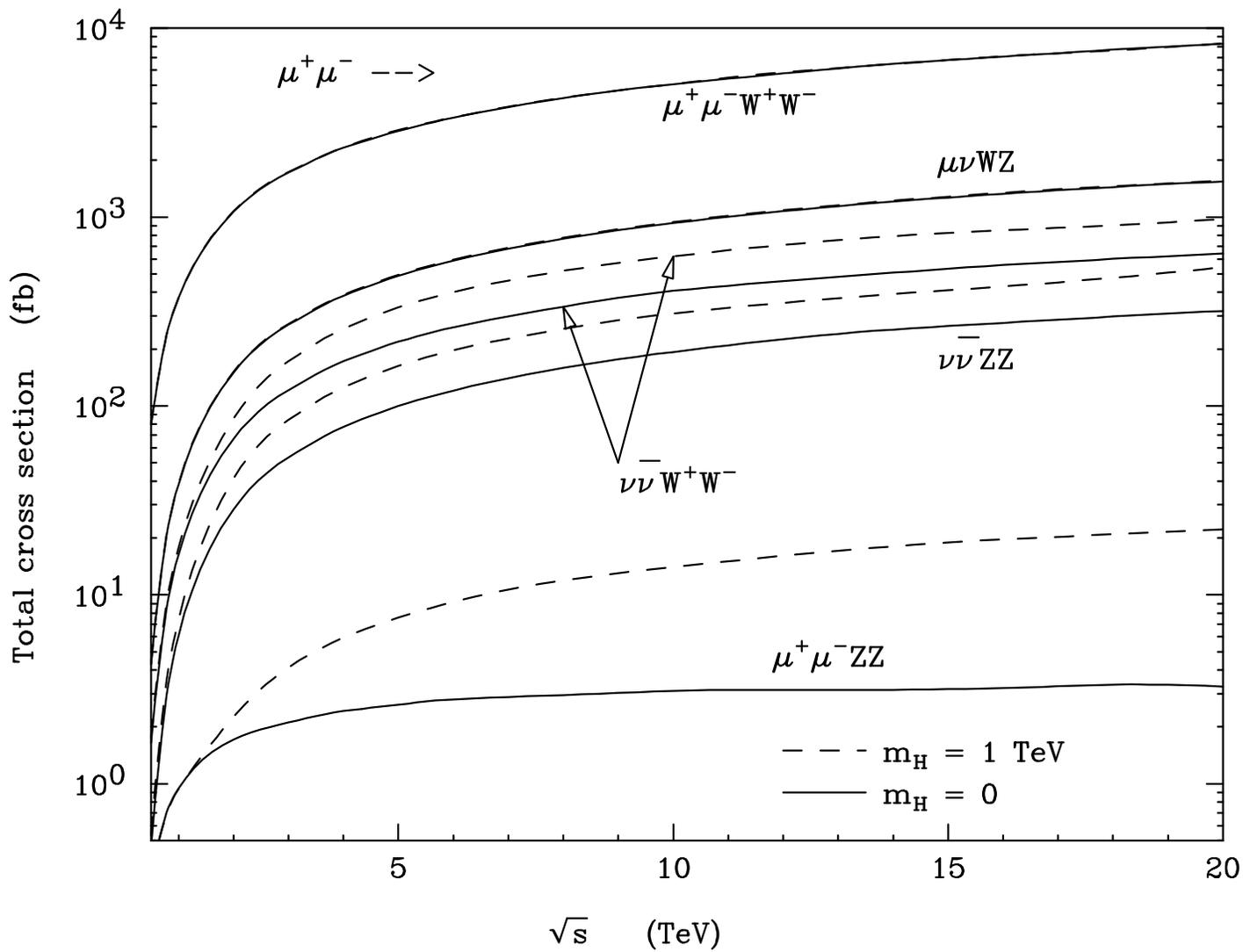
\end{document}